\documentclass[twocolumn,preprintnumbers, amssymb,amsmath,aps,prd,floatfix,nofootinbib,superscriptaddress,showpacs]{revtex4-1}

\usepackage{epsfig}
\usepackage{bm}
\usepackage{amssymb}
\usepackage{amsmath}
\usepackage{color}
\usepackage{subfigure}
\usepackage[colorlinks,
            linkcolor=blue,
            anchorcolor=black,
            citecolor=blue
            ]{hyperref}

\usepackage{soul}

\newcommand{\nn}{\nonumber \\}

\begin{document}

\title{Understanding Gravitational Form Factors with the Weizs\"acker-Williams Method}

\author{Yoshikazu Hagiwara}
\email{hagiyoshihep@gmail.com} 
\affiliation{School of Science and Engineering, The Chinese University of Hong Kong, Shenzhen, Guangdong, 518172, P.R. China}
%\affiliation{School of Science and Engineering, The Chinese University of Hong Kong, Shenzhen (CUHK-Shenzhen), Guangdong, 518172, P.R. China}
\affiliation{University of Science and Technology of China, Hefei, Anhui, 230026, P.R.China}

\author{Xuan-Bo Tong}\email{xuan.bo.tong@jyu.fi} 
\affiliation{Department of Physics, University of Jyväskylä, P.O. Box 35, 40014 University of Jyväskylä, Finland}
\affiliation{Helsinki Institute of Physics, P.O. Box 64, 00014 University of Helsinki, Finland}
\affiliation{School of Science and Engineering, The Chinese University of Hong Kong, Shenzhen, Guangdong, 518172, P.R. China}

\author{Bo-Wen Xiao}\email{xiaobowen@cuhk.edu.cn} 
\affiliation{School of Science and Engineering, The Chinese University of Hong Kong, Shenzhen, Guangdong, 518172, P.R. China}
\affiliation{Southern Center for Nuclear-Science Theory (SCNT), Institute of Modern Physics, Chinese Academy of Sciences, Huizhou 516000, Guangdong Province, China}

\begin{abstract}
Understanding the internal structure of nucleons and nuclei has been a topic of enduring interest in high-energy physics. Gravitational form factors~(GFFs) provide an important portal for us to probe the energy-momentum/mass distribution of nucleons and nuclei. This letter presents the study of the photon and gluon momentum GFFs, also known as the A-GFFs, of relativistic hadrons using the Weizs\"acker-Williams method. To begin, we express the photon A-GFFs in terms of charge form factors and discuss the corresponding photon radius. Furthermore, an integral relation between the gluon A-GFF and the Laplacian of dipole scattering amplitude is derived in the small-$x$ framework, and it allows us to unravel the gluon energy momentum distribution inside hadrons through measurements at the upcoming Electron-Ion Collider. In addition, we generalize the analysis to study the A-GFF of nuclei and propose employing the nuclear gluon mean square radius, together with the charge distribution, to constrain the neutron distribution for large nuclei. This work provides an interesting perspective into the fundamental structure of high-energy hadrons. 
\end{abstract}
\maketitle
\section{Introduction}
Gravitational form factors (GFFs)~\cite{Kobzarev:1962wt,Pagels:1966zza,Ji:1996ek}, defined as the transition matrix elements of energy-momentum tensor (EMT), encode profound fundamental information of hadrons~(see a recent review~\cite{Burkert:2023wzr}), such as the distribution of the hadron mass~\cite{Ji:1994av,Ji:1995sv,Ji:2021mtz,Hatta:2018sqd,Lorce:2017xzd,Metz:2020vxd,Lorce:2021xku,Yang:2018nqn} and spin~\cite{Jaffe:1989jz,Ji:1996ek,Leader:2013jra,Wakamatsu:2014zza,Ji:2020ena,Ji:2020hii} as well as the nature of the internal stress~\cite{Polyakov:2002yz,Polyakov:2018zvc,Lorce:2018egm,Freese:2021czn}. Constraints on hadronic GFFs can be accessed through exclusive processes such as the deeply virtual Compton scattering~(DVCS)~\cite{Ji:1996ek,Ji:1996nm,Pasquini:2014vua,Burkert:2018bqq,Dutrieux:2021nlz,Burkert:2021ith,Burkert:2023atx,Bhattacharya:2022xxw,Bhattacharya:2023wvy} and near-threshold heavy-quarkonium productions~\cite{Kharzeev:1995ij,Kharzeev:1998bz,Hatta:2018ina,Hatta:2019lxo,Kharzeev:2021qkd,Boussarie:2020vmu,Hatta:2021can,Guo:2021ibg,Guo:2023qgu,Guo:2023pqw,Sun:2021gmi,Sun:2021pyw,Mamo:2021tzd,Mamo:2022eui,Wang:2021dis,Duran:2022xag} at Jefferson Laboratory~\cite{Dudek:2012vr,Accardi:2023chb}, the electron-ion collider~(EIC)~\cite{Boer:2011fh,Accardi:2012qut,AbdulKhalek:2021gbh,Anderle:2021wcy}, the large hadron electron collider~\cite{Hatta:2023fqc,LHeC:2020van}, and also ultraperipheral heavy-ion collisions. Lattice simulations~\cite{Hagler:2007xi,Deka:2013zha,Shanahan:2018nnv,Shanahan:2018pib,Hackett:2023nkr,Hackett:2023rif} and perturbative analysis~\cite{Hatta:2018sqd,Tanaka:2022wzy,Tong:2021ctu,Tong:2022zax,Anikin:2019kwi,Dehghan:2023ytx} have provided a lot of theoretical insights into the understanding of hadronic GFFs~\cite{Burkert:2023wzr}.

Of particular interest is the study of GFFs in fast-moving hadrons~\cite{Burkardt:2000za,Burkardt:2002hr,Lorce:2018egm,Freese:2021czn,Freese:2019bhb,Freese:2022fat,Panteleeva:2021iip,Kim:2021jjf,Freese:2021mzg,Panteleeva:2022uii,Panteleeva:2023evj,Freese:2023abr,Choudhary:2022den,Strikman:2003gz}. First, the Lorentz contraction facilitates a probabilistic interpretation of spatial transverse parton densities and associated GFFs, allowing for physical radii to characterize the intrinsic size of hadrons~(see the discussions in e.g., \cite{Burkardt:2000za,Freese:2021czn}). Furthermore, the Wigner distributions~\cite{Ji:2003ak,Belitsky:2003nz} and generalized transverse momentum distributions~(GTMDs)~\cite{Meissner:2008ay,Meissner:2009ww,Lorce:2013pza} provide the full quantum information stored in a quantum system. Currently, the Weizsäcker-Williams (WW) method~\cite{Fermi:1924tc,vonWeizsacker:1934nji,Williams:1935dka}, as a powerful and widely-used tool in computing these distributions~\cite{McLerran:1993ni,McLerran:1993ka,McLerran:1994vd,Kovchegov:1996ty,Kovchegov:1997pc,Jalilian-Marian:1996mkd,Kovchegov:1998bi,Mueller:1999wm,Dominguez:2010xd,Dominguez:2011wm,Hatta:2017cte,Hatta:2016dxp,Hagiwara:2017fye,Boussarie:2018zwg,Pasechnik:2023mdd}, remains relatively unexplored in the GFF studies.

The WW method, also known as the equivalent photon approximation~\cite{Fermi:1924tc}, has been successful in describing the soft photon radiations from relativistic charged particles in ultraperipheral nuclei collisions~\cite{Bertulani:1987tz,Krauss:1997vr,Bertulani:2005ru,Baltz:2007kq}. Its application to QCD allows us to define and compute the WW small-$x$ gluon distributions radiated from color charges~\cite{McLerran:1993ni,McLerran:1993ka,McLerran:1994vd,Kovchegov:1996ty,Kovchegov:1997pc,Jalilian-Marian:1996mkd,Kovchegov:1998bi,Mueller:1999wm,Dominguez:2010xd,Dominguez:2011wm}. The WW gluon distribution, as an essential ingredient in the color-glass-condensate~(CGC) formalism, is vital to describe the gluon saturation phenomena~\cite{Gribov:1983ivg,Mueller:1985wy,Mueller:1989st,Iancu:2003xm,Gelis:2010nm,Kovchegov:2012mbw,Morreale:2021pnn}. Especially, the small-$x$ WW gluon GTMDs have been under intensive study recently~\cite{Hatta:2016dxp,Hagiwara:2017fye,Boussarie:2018zwg,Pasechnik:2023mdd}. Given their close connections to GFFs, it is natural to also use the WW method to compute the gluon GFFs and the associated radii in the small-$x$ formalism. 

The objective of this paper is to use the WW method to study the photon and gluon GFFs of high-energy hadrons. As a first attempt, we focus on the A-GFFs, pivotal in determining the light-front momentum radius \cite{Freese:2021czn}. First, we compute the photon GFFs as a simple example and proof of principle. Then, through the small-$x$ gluon GTMDs, we relate the gluon A-GFFs to the impact parameter dependent color-dipole scattering amplitude. Our results reveal that the gluon GFFs and radius are sensitive to the saturation momentum of hadron systems, and help in understanding the geometry and mass distribution of proton and nuclei. The extension to other GFFs will be addressed in a future study.

\section{The photon A-GFFs and radii}
\subsection{The photon A-GFFs in WW method}

Let us first consider the photon GFFs, for which the associated EMT is defined as follows:
\begin{align}
T^{\mu\nu}_\gamma =  F^{\mu\alpha}F_{\alpha}^{~\nu} + \frac{g^{\mu\nu}}{4} F^{\alpha\beta}F_{\alpha\beta}~,
\label{eq:defEMT}
\end{align} 
where $F^{\mu\alpha}$ represents the photon field strength tensor. As a first step, we assume the target charged hadron is spin-0. The photon GFFs for such a hadron can be parametrized as~\cite{Polyakov:2018zvc}:
\begin{align}
\langle p'| T^{\mu\nu}_\gamma|p\rangle = & 2 P^\mu P^\nu A_\gamma(t)
+ C_\gamma(t)\frac{\Delta^\mu \Delta^\nu - g^{\mu\nu}\Delta^2}{2} 
\notag \\ 
& + 2m^2 \bar{C}_\gamma(t) g^{\mu\nu}~,
\label{eq:defGFF}
\end{align}
where $p=P- \Delta/2$ and $p'=P+ \Delta/2$ are the initial and final state hadron momentum, respectively. $P$ is the average momentum, $\Delta$ is the momentum transfer with $t=\Delta^2$, and $m$ denotes the hadron mass. The above photon GFFs, defined as the matrix elements of the EMT operator, provide the vital information on the energy momentum distribution from photons. Specifically, the $A_\gamma(t)$ and $C_\gamma(t)$ GFFs encode information about mass/momentum and shear pressure distribution, respectively, while $\bar{C}_\gamma(t)$ arises from the non-conservation of the photon EMT. In this paper, we mainly focus on the $A(t)$ GFF and comment on the possible extension to other GFFs at the end.

\begin{figure}[htbp]
\centering
\includegraphics[scale=0.65]{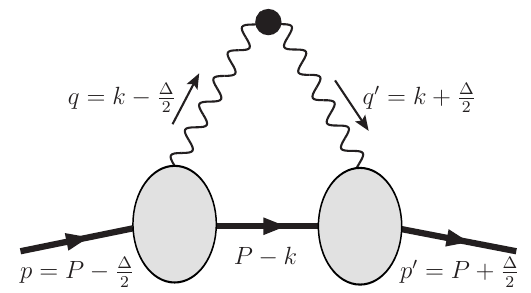}
\caption{The LO diagram contributing to the photon GFFs. The grey bubbles stand for the electromagnetic currents of the hadron and the back dot on the top represents the vertex of the photon EMT operators.}  
\label{fig:QED}
\end{figure}

In collider experiments, charged particles often travel relativistically, where the WW method applies~\cite{Fermi:1924tc,vonWeizsacker:1934nji,Williams:1935dka}. In the relativistic regime, perturbed by a small transverse momentum transfer $|t|= \Delta^2_\perp$, the initial and final-state hadrons in Eq.~(\ref{eq:defGFF}) travel rapidly along nearly the same direction. According to the WW method, the electromagnetic field of the fast-moving hadron can be treated as a swarm of quasi real photons in the transverse plane. These photons originate from the rapid motion of the hadronic charged constituents, namely, the current $J^\mu=e\bar \psi \gamma^\mu \psi$. Consequently, the computation of photon GFFs becomes the evaluations of photon bremsstrahlung from the hadron and the interaction between the photon and the EMT operator. Fig.~\ref{fig:QED} presents the leading-order-$\alpha$ diagram of photon GFFs, where two photon exchanges are necessitated by the coupling between the EMT and the charged currents. In the following, we will show that the photon A-GFF can be expressed in terms of the electromagnetic (EM) form factor of the relativistic hadrons. We are interested in the leading-power contribution in the limit $P^+\gg | \Delta_\perp|,m$ and work in the leading order of $\alpha$.

%Chapter 11

%The WW method allows us to compute the distribution of quasi-real photon radiated from a fast-moving charged hardon. 
%Before the detailed analysis, it is useful to recall two essential approximations employed by the Weizsäcker-Williams method in high-energy scatterings: (i) The fast-moving hadron is approximated as a classical charged source that generates photon radiations. (ii) The limit of small photon virtuality is taken,  $q^2 \rightarrow 0$, capturing the most significant contributions from these photon radiations. %Second, the dominant contributions of the photon radiations arise from regions where the photon virtuality is minimal, $q^2 \rightarrow 0$. %These approximations will be justified further in our subsequent calculation of the photon A-GFF.

%Applying the Weizs\"acker-Williams method in computing the photon radiations from a relativistic hadron includes the following two approximations: (i) The fast-moving hadron is treated as a classical charged source that generates the photon radiations. (ii) The quasi-real photon approximation due to small photon virtuality is taken, capturing the most significant contributions from these photon radiations.
 To facilitate the detailed analysis, we work in the infinite momentum frame: $\Delta^\mu=\big(0, 0, \Delta_\perp\big)$, $P^\mu=\big( P^+,\frac{ \Delta^2_\perp/4}{2 P^+}, 0\big)$, we can find the A-GFF directly from the longitudinal component of the EMT matrix, 
\begin{align}
A_\gamma(t)=\frac{1}{2( P^+)^2}\langle p'| T^{++}_\gamma|p\rangle~. 
\label{eq:newpro}
\end{align}
Using the QED perturbation theory, the leading-$\alpha$ contribution to the EMT matrix in Fig.~\ref{fig:QED} can be expressed as
\begin{align} 
\langle p'|&T^{++}_\gamma|p\rangle
\notag \\   =&
4\pi \alpha  \int 
\frac{d^4 q~d^4 q' }{(2\pi)^8} \frac{i^2 (q^{+}g^{\lambda}_{\perp \sigma} -q^{\lambda}_\perp g^{+}_{\ \sigma} ) ( q'^{+} g^\perp_{\lambda\rho}-q'_{\perp\lambda} g^{+}_{\ \rho} ) }{(q^2+ i\epsilon)(q'^2+ i\epsilon)}  
\notag \\ &
\times \int d^4 x d^4 ye^{iq\cdot x-iq'\cdot y} \langle p' | {\cal T} \{J^\sigma(x) J^\rho(y) \}~|p\rangle ~,
\label{appeq:GFF1}
\end{align}
where the two photons carry the momenta $q$ and $q'$, respectively, with their propagators specified in the Feynman gauge for convenience.

In Eq.~(\ref{appeq:GFF1}), the charge source of the two photon radiations is effectively described by the correlation matrix of the two electromagnetic current $J^\mu(x)=\bar \psi(x)\gamma^\mu \psi(x)$.  %Here, we implied the summation over quark flavors in $J^\mu$. 
Then, we insert a complete set of particle states between the two currents using $\sum_N |N\rangle \langle N |=\mathbf{1}$. Theoretically speaking, arbitrary number of particle states $|N\rangle$ may serve as intermediate states in mediating the two photon exchanges. However, given the small momentum transfer, the probability of fragmentation or decay during this forward scattering is exceedingly low. Consequently, the predominant contributions from intermediate states are most likely from the elastic channel, where the fast-moving hadron system remains intact throughout the process. This dominance of the elastic channel in the photon A-GFF indicates that the fast-moving hadron can be viewed as a classical charge source. 

Thus, the charge source in Eq.~(\ref{appeq:GFF1}) at leading power can be written as 
\begin{align}
 \langle p' | &{\cal T} [J^\sigma(x) J^\rho(y) ]
|p\rangle
=\int\frac{d^4  \ell }{(2\pi)^4 }(2\pi)\delta(\ell^2-m^2)\theta(\ell^+)
\notag \\&\times\bigg[
\langle p' | J^\sigma(x)|\ell \rangle \langle \ell  | J^\rho(y)
|p\rangle  
+ \langle p' |  J^\rho(y)|\ell \rangle \langle \ell  |J^\sigma(x) 
|p\rangle  \bigg]
\notag \\&+{\cal O}\Big(\frac{\Delta_\perp}{P^+}, \frac{m}{P^+}\Big)~.
\label{appeq:GFF2}
\end{align} 
The EM form factors will become explicit after the parametrization of the current matrices. 
For a spinless hadron or a large nuclei, the EM form factors is given by 
\begin{align}
\langle p_2|J^\mu(x)|p_1 \rangle=e^{i(p_2-p_1 )\cdot x}e(p_1+p_2)^\mu F((p_1-p_2)^2)~,
\label{appeq:GFF3}
\end{align}
where $F$ represents the charge form factor of the hadron, normalized such that $F(0)=Z$, with $Z$ being the charge number. Using Eqs.~({\ref{appeq:GFF1},\ref{appeq:GFF2},\ref{appeq:GFF3}), the photon A-GFF can be expressed in terms of the charge form factors: 
\begin{align}
 A_\gamma(t)=&8 \alpha 
\int\frac{d^4 \ell }{(2\pi)^2}\delta(\ell^2-m^2)\theta(\ell^+) q_\perp \cdot q'_\perp 
 \frac{ F(q^2)}{q^2} \frac{ F(q'^2)}{q'^2} 
\end{align}
where the photon momenta become $q=p-\ell$ and $q'=p'-\ell$. It is useful to change the variables from $\ell$ to $k$ through $k=P-\ell$ with $k^+=xP^+$.  The variable $x$ represents the longitudinal momentum fraction of the radiated photons, and the transverse momenta are given by $q_\perp=k_\perp-\bar x\Delta_\perp/2$ and $q'_\perp=k_\perp+\bar x\Delta_\perp/2$ with $\bar x\equiv 1-x$. Using the on-shell condition $\ell^2=m^2$, the photon virtualities are given by
\begin{align}
q^2=-\big(q_\perp^2+x^2m^2\big)/\bar x, \quad q'^2=-\big(q'^2_\perp+x^2m^2 \big)/\bar x~.
\end{align}
After this change of variables, we obtain
\begin{align}
 &A_\gamma(t) 
 \notag \\
=&4 \alpha \int^{1}_0dx \int\frac{ d^2 k_\perp}{(2\pi)^2}
\bar x\Big(k_\perp +\bar x\frac{ \Delta_\perp}{2}\Big)\cdot \Big(k_\perp -\bar x\frac{  \Delta_\perp }{2}\Big)
\notag \\ & \times
\frac{F(q'^2)F(q^2)
}{\big[(k_\perp+\bar x\frac{ \Delta_\perp}{2})^2+m^2x ^2\big]\big[(k_\perp -\bar x
\frac{\Delta_\perp}{2} )^2+m^2x^2\big]}~.
\label{appeq:GFF4}
\end{align} 
Given that the hadronic charge form factors decrease rapidly at large momentum transfers~\cite{Efremov:1979qk}, the $k_\perp$-integration in Eq.~(\ref{appeq:GFF4}) converges in the ultraviolet region for a hadron target.
%Eq.~(\ref{appeq:GFF4}) illustrates the dominance of soft or quasi-real photons, 

 Furthermore, the dominant contributions to the integral in Eq.~(\ref{appeq:GFF4}) originate from the region where the denominator of the photon propagator, $q^2$, approaches zero. In this region, the radiated photons are quasi-real with $k_\perp\sim xm \sim \Delta_\perp\ll P^+$. This validates the small-$x$ limit. Finally, we obtain the photon A-GFF in the WW approximation: 
\begin{align}
A_\gamma(t)
=&4 \alpha \int^{1}_0dx \int\frac{ d^2 k_\perp}{(2\pi)^2}
\frac{q_\perp\cdot q_\perp'}{\big(q_\perp^2+m^2x ^2\big)\big(q'^2_\perp+m^2x^2\big)}
\notag \\ &
F\big[-(q_\perp^2+x^2m^2\big)]F\big[-(q'^2_\perp+x^2m^2)\big]
\label{eq:AGFFs}~,
\end{align}
where $q_\perp\approx k_\perp-\Delta_\perp/2$ and $q'_\perp\approx  k_\perp+\Delta_\perp/2$.

% Then we calculate the A-GFF in Fig.~\ref{fig:QED} by putting the one-particle intermediate state on shell and cast the photon A-GFF into
  %\begin{align}
   %  A_\gamma(t) &= 4 \alpha \int_0^1   d x \int \frac{d^2   k_\perp}{(2\pi)^2} \frac{ q_\perp\cdot  q'_\perp}{q^2q'^2}F(q^2)F(q'^2)~.
  %   \label{eq:AGFFs}
%\end{align} 
%Here $F$ represents the charge form factor of a spinless hadron, parametrized from 
%$\langle p_2|J^\mu|p_1 \rangle=e(p_1+p_2)^\mu F((p_1-p_2)^2)
%$. We introduce the variable $  q_\perp= k_\perp-  \Delta_\perp/2$, $  q^\prime_\perp= k_\perp+ \Delta_\perp/2$ for convenience, and $q^2=-( q^2_\perp+x^2 m^2)$ represents the virtuality of the photons, with $x$ as the longitudinal momentum fraction. In evaluating Eq.~(\ref{eq:AGFFs}), we have taken the small-$x$ limit in the integrand, as the photon A-GFF is dominated by the soft photon with small virtuality. In this region, since $xm \sim  | k_\perp|\ll P^+$, we retain the $xm$-terms in $q^2$ and $q^{\prime 2}$. 

Alternatively, the photon A-GFF in the WW approximation can be obtained from the small-$x$ WW photon GTMDs~\cite{Xiao:2020ddm,Klein:2020jom} $xf_\gamma(x, k_\perp, \Delta_\perp)$ as follows
\begin{equation}
A_\gamma(t)=\int_0^1 d x\int d^2  k_\perp~xf_\gamma(x, k_\perp, \Delta_\perp)~.
\label{eq:relation}
\end{equation}
One can show that the above result leads to the GFF consistent with that in Eq.~(\ref{eq:AGFFs}). 

 \subsection{The photon radius}
Of particular interest is the small-$t$ behavior of the A-GFF, as it determines various energy momentum radii~\cite{Polyakov:2002yz,Polyakov:2018zvc,Freese:2021czn}~(see a recent review in~\cite{Burkert:2023wzr}). For a fast-moving particle, we can define the photon light-front momentum density in the impact parameter $b_\perp$ space, through the two-dimensional~(2D) inverse Fourier transform~\cite{Lorce:2018egm,Freese:2021czn}:
\begin{align}
\epsilon_\gamma( b_\perp)=
\int \frac{d^2 \Delta_\perp}{(2\pi)^2} e^{-i  \Delta_\perp \cdot  b_\perp}
A(- \Delta^2_\perp)~.
\label{eq:ep}
\end{align}
The associated mean-square transverse radius can then be computed by the slope of the $A$-GFF at $t=0$~\cite{Freese:2021czn,Freese:2019bhb}:
\begin{align}
\langle  b^2_\perp\rangle_\gamma\equiv\frac{ \int d^2 b_\perp b^2_\perp\epsilon_\gamma( b_\perp)}{\int d^2 b_\perp\epsilon_\gamma( b_\perp)}=\frac{4}{A_\gamma(0)} \frac{dA_\gamma(t)}{d t} \Big \vert _{t=0}~.
\label{Eq:def}
\end{align}
%In these statements, we first demonstrate
Interestingly, the photon momentum radius from a charged particle is singular due to the long-range tail of the Coulomb field. To see this, let us first consider a pointlike charge, where $F(q^2)=1$. The asymptotic behavior of photon A-GFF at small-$|t|$ can be computed as follows: 
\begin{align}
A_\gamma(t)=\frac{\alpha}{\pi} \left[\text{U.V.}+\frac{t}{m^2}\Big(\frac{3 }{16}\frac{m\pi^2 }{\sqrt{-t}} -\frac{1}{3}\Big)+\cdots\right]~,
\label{eq:Asmallt}
\end{align}   
where $\text{U.V.}$ represents the ultraviolet divergent part due to the point-like charge input and one can find $\text{U.V.}=\frac{1}{\epsilon}+\ln \frac{\mu^2}{m^2}+1$ in the dimensional regularization. The rest part of the A-GFF is finite at $t=0$. However, due to the presence of a non-analytic term in Eq.~(\ref{eq:Asmallt}), the slope $dA_\gamma/d t$ diverges at $t=0$, resulting in an infinite photon momentum radius. Since the small-$t$ behavior of $A_\gamma(t)$ corresponds to the large-$b_\perp$ behavior of the light-front momentum density $\epsilon_\gamma(b_\perp)$, the infinite photon momentum radius $\langle b_\perp^2 \rangle_\gamma$ reflects the fact that the energy in the Coulomb field extends out to infinity. This non-analytic singularity is inherent to any charged particle, irrespective of the internal structure, due to the long-range nature~\cite{Donoghue:2001qc,Varma:2020crx,Freese:2022jlu}.%, as it arises from the long-range tail of the Coulomb field
%In fact, this non-analytic singularity is inherent to any charged particle, irrespective of the internal structure, as it arises from the long-range tail of the Coulomb field
%\begin{align}
 %\epsilon_\gamma({b}_\perp)
%= &\frac{ \alpha}{\pi^2} \int_0^1 d x m^2 x^2\left[K_1\big(m x{b}_\perp\big)
%\right]^2,
%\end{align}
%\begin{align}
%\epsilon_\gamma(b_\perp)\sim \frac{1}{b_\perp^2}
%\end{align}  
\begin{figure}[htbp]
\centering
\includegraphics[scale=0.45]{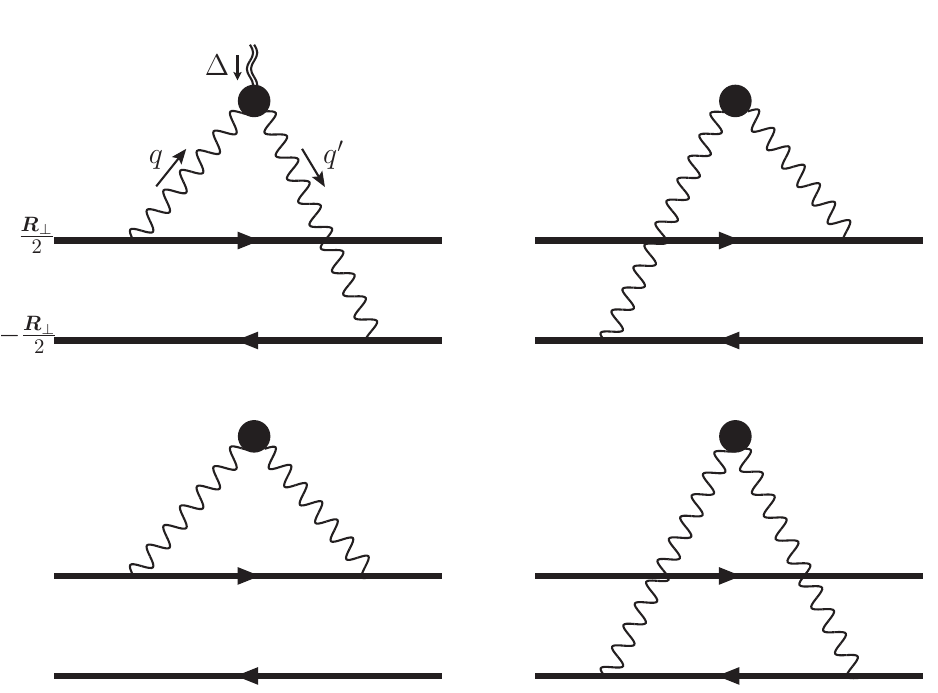}
\caption{Photon GFFs in a charge neutral dipole. } 
\label{fig:dipole}
\end{figure} 

However, for charge-neutral particles, this singular behavior cancels out due to internal charge correlations.  To illustrate this cancellation, we can consider a scalar dipole model by pairing the positive and negative charged particles at a distance of $ R_\perp$~\cite{Mueller:1993rr,Mueller:1999yb,Kovchegov:2012mbw}, as shown in Fig.~\ref{fig:dipole}. Summing up all four contributions gives 
\begin{align}
A_\gamma(&t)
=4\alpha \int_0^1  d x \int \frac{d^2   q_\perp d^2   q'_\perp}{(2\pi)^4} 
\frac{ q_\perp\cdot  q'_\perp}{( q^2_\perp+m^2x^2)( q'^2_\perp+m^2x^2)} 
\notag \\ 
& \times
(2\pi)^2\delta^{(2)}( \Delta_\perp+ q_\perp- q'_\perp)
\notag \\ 
& \times \big[e^{-i  q_\perp \cdot   R_\perp/2}
-e^{i  q_\perp\cdot   R_\perp/2}\big]F(-( q^2_\perp+m^2x^2))
\notag \\ 
&\times \big[e^{i  q'_\perp \cdot   R_\perp/2}
-e^{-i  q'_\perp\cdot   R_\perp/2}\big]F(-( q'^2_\perp+m^2x^2))~,
\label{eq:dipoleAGFF}
\end{align} 
where the phase factors account for the relative positions of the charges. %One can find that the opposite charges remove the singular behavior due to the long-range Coulomb fields, allowing for a finite photon momentum radius.

For the long-range behavior, it suffices to consider the dipole composed of pointlike charges. The associated light-front momentum density is given by: 
\begin{align}
 \epsilon_\gamma({b}_\perp)
= &\frac{ \alpha}{\pi^2} \int_0^1 d x m^2 x^2\left[\frac{{b}_\perp+{R}_\perp / 2}{\left|{b}_\perp+{R}_\perp / 2\right|} K_1\big(m x\left|{b}_\perp+{R}_\perp / 2\right|\big)\right.
\notag \\ &\left.\quad-\frac{{b}_\perp-{R}_\perp / 2}{\left|{b}_\perp-{R}_\perp / 2\right|} K_1\big(m x\left|{b}_\perp-{R}_\perp / 2\right|\big)\right]^2,
\end{align}
where the function $K_1$ refers to the modified Bessel function of the second kind of order one. In the limit $b_\perp \gg R_\perp$, we can derive its contribution to the photon radius $\langle b_\perp^2 \rangle_\gamma $:
\begin{align} 4\frac{d A_\gamma(t)}{dt} \Big\vert_{t=0}=&\int d^2b_\perp {b}_\perp^2 \epsilon_\gamma({b}_\perp)\Big\vert_{b_\perp \gg R_\perp }
\notag \\ 
\approx& \int_0^1 d x \frac{ \alpha}{2 m^2 x^2}\left[I_0(m x|{R}_\perp|)-1\right]
\notag \\ 
 \simeq& \frac{\alpha R_\perp^2}{8}~,
\label{eq:sing}
\end{align}
where the function $I_0$ stands for the modified Bessel function of the first kind of order zero, and and the small-$x$ approximation has been applied. In this case, the long-range singularitie clearly cancel out, and the resulting photon radius is finite.

\subsection{Extention to spin-1/2 hadrons}
The analysis for spin-0 hadrons can be extended to spin-1/2 particles.  For these particles, the photon A-GFF in the WW approximation can be expressed in terms of the associated EM form factors as follows
\begin{align}
A_\gamma&(t)=4 \alpha\int^1_0 d x\int\frac{  d^2 k_\perp}{ (2\pi)^2  }
\frac{ q_\perp \cdot   q'_\perp}{( q_\perp^2+m^2x^2)
(q'^2_\perp+m^2x^2) }   
\notag \\ 
&\times\bigg\{ 
F_1(q^2)F_1(q'^2) +  \frac{F_2(q^2)F_2(q'^2)}{4m^2} q_\perp \cdot q'_\perp\bigg\}~,
\label{appeq:nGFF1}
\end{align} 
where $F_1$ is the Dirac form factor, and $F_2$ is the Pauli form factor
\begin{align}
&\langle p_2,\lambda'|J^\mu| p_1,\lambda\rangle=\bar u_{\lambda'} (p_2)\Big[F_1((p_2-p_1)^2) \gamma^\mu 
\notag \\
&\quad\quad\quad+F_2((p_2-p_1)^2) \frac{i \sigma^{\mu \alpha} (p_2-p_1)_\alpha}{2 m}\Big] u_\lambda(p_1)~,
%\label{appeq:nEM}
\end{align}
where $u_\lambda(p)$ is the Dirac spinor of the hadron with momentum $p$ and helicity $\lambda$.

 Since the A-GFFs arise from the $T^{++}$-component, which account for unpolarized effects in high-energy hadrons with spin, the primary difference in analyzing spin-$1/2$ hadrons compared to spin-0 hadrons lies in averaging over the hadron's polarizations. Thus, the projection in Eq.~(\ref{eq:newpro}) becomes 
\begin{align}
A_\gamma(t) =\frac{1}{2}\sum_{\lambda=\uparrow,\downarrow}\frac{1}{2(P^{+})^2}\left\langle p^{\prime}, \lambda\left|T_\gamma^{++}\right| p, \lambda\right\rangle~.
\label{appeq:nGFFprojection}
\end{align} 
Here, only the helicity-conserved EMT amplitude is included, as it provides the dominant contribution.

Furthermore, when applying the WW approximation in the current correlation matrix as Eq.~(\ref{appeq:GFF2}), the polarization sum in the intermediate state must also be considered:
  \begin{align}
 \langle p',\lambda |&  {\cal T} [J^\sigma(x) J^\rho(y) ]
|p,\lambda \rangle
\notag \\ 
=& \sum_{s=\uparrow,\downarrow}\int \frac{d \ell^+ d^2\ell_\perp}{2\ell^+(2\pi)^3}\Big[
\langle p',\lambda| J^\sigma(x)|\ell,s \rangle \langle \ell,s  | J^\rho(y)
|p,\lambda\rangle 
\notag \\ &
+ \langle p',\lambda|J^\rho(y)|\ell,s \rangle \langle \ell,s  |  J^\sigma(x)
|p,\lambda\rangle\Big]~.
\label{appeq:nGFF2}
\end{align} 
Applying this correlation matrix in Eq.~(\ref{appeq:GFF1}) leads to
\begin{align} 
&\langle p',\lambda|T^{++}_\gamma|p,\lambda\rangle \notag \\ 
=& 8
\pi \alpha  \int \frac{d \ell^+ d^2\ell_\perp}{2\ell^+(2\pi)^3}
 \frac{i^2 (q^{+}g^{\lambda}_{\perp \sigma} -q^{\lambda}_\perp g^{+}_{\ \sigma} ) ( q'^{+} g^\perp_{\lambda\rho}-q'_{\perp\lambda} g^{+}_{\ \rho} ) }{(q^2+ i\epsilon)(q'^2+ i\epsilon)}
\notag \\ 
&\times   
\bigg\{F_1\left(q^2\right) F_1\left(q'^2\right) \bar{u}_\lambda\left(p'\right)\gamma^\sigma u_\lambda(\ell)\bar u_\lambda(\ell)
 \gamma^\rho u_\lambda(p)
\notag\\
&~~~~~~+F_2(q^2)F_2(q'^2)
\bar{u}_\lambda\left(p'\right) \frac{i \sigma^{\sigma \alpha} q'_\alpha}{2 m} u_{-\lambda}(\ell)
\notag\\
&
~~~~~~\times\bar u_{-\lambda}(\ell)  \frac{i \sigma^{\rho \alpha} (-q)_\alpha}{2 m}u_\lambda(p)
\bigg\}~,
\label{appeq:nGFF3}
\end{align}
 where $q=p-\ell$ and $q'=p'-\ell$. As in previous analyses, it is found that this photon EMT amplitude is dominated by soft photon radiation, allowing the small-$x$ limit to be applied. By evaluating the Dirac structures and using Eq.~(\ref{appeq:nGFFprojection}), the photon A-GFF of the spin-$1/2$ hadron can be extracted, yielding the result given in Eq.~(\ref{appeq:nGFF1}).

%Furthermore, because the helicity-flip amplitudes are known to be power-suppressed in the relativistic limit~($\bar{u}_{\lambda}\left(p^{\prime}\right) \gamma^{+} u_{-\lambda}(p)\approx 0$), one can alternatively utilize the spin-averaged amplitude to do the projection:This indicates that the A-GFFs account for the unpolarized effects in a high-energy hadron with spin.

\iffalse
In fact, because the helicity-flip amplitudes are power-suppressed in the relativistic limit~($\bar{u}_{\lambda}\left(p^{\prime}\right) \gamma^{+} u_{-\lambda}(p)\approx 0$), one can alternatively utilize the spin-averaged amplitude to do the projection:
\begin{align}
A_\gamma(t) \Big \vert_{spin-1/2}=&\frac{1}{2}\sum_{\lambda=\uparrow,\downarrow}\frac{1}{2(P^{+})^2}\left\langle p^{\prime}, \lambda\left|T_\gamma^{++}\right| p, \lambda\right\rangle
\notag \\ 
\approx&\frac{1}{2}\sum_{\lambda,\lambda'=\uparrow,\downarrow}\frac{1}{2(P^{+})^2}\left\langle p^{\prime}, \lambda'\left|T_\gamma^{++}\right| p, \lambda\right\rangle~.
\label{eq:A}
\end{align}
This indicates that the A-GFFs account for the unpolarized effects in a high-energy hadron with spin.
\fi
\par 
\section{Gluon GFFs and radii}  
Now let us generalize our analysis to the gluon GFFs of relativistic hadrons. In CGC, the small-$x$ gluons dominate the parton distribution at high energy. Similar to the QED case outlined in Eq.~(\ref{eq:relation}), we can obtain the relation between the gluon A-GFF and the unpolarized gluon GTMDs
\begin{align}
A_g(t)=\int_0^1 d x\int d^2k_\perp x{\cal G}_x( k_\perp, \Delta_\perp)~,
\end{align}which is in agreement with previous results regarding the gluon Generalized Parton Distributions and GFFs at zero skewness~\cite{Ji:1998pc}. 

There are two types of GTMDs in the CGC formalism, namely, the dipole and WW gluon GTMDs. As shown in Ref.~\cite{Hatta:2016dxp}, these GTMDs are related to the Fourier transforms of color dipole and quadrupole correlators, respectively. %For clarity and to aid the reader's understanding of gluon GTMDs, we define each type explicitly. %One can write them as follows
%\begin{align}
%	x\mathcal{G}_{x}^\text{WW}(k_\perp,&\Delta_\perp) =\frac{2N_c}{\alpha_s} \int \frac{d^2 b_{1} d^2 b_{2}}{(2\pi)^4} e^{-i\Delta_\perp \cdot\frac{b_1+ b_2}{2} -ik_\perp\cdot(b_1-b_2)} \notag \\
% &\times\frac{1}{N_c} \big \langle \text{tr}\big[(i\partial^j_\perp U_{b_1}) U^\dagger_{b_2} ( i\partial^j_\perp U_{b_2}) U_{b_1}^\dagger \big]\big \rangle_x~, \\
%	x\mathcal{G}_x^{\text{DP}}(k_\perp,&\Delta_\perp)=
%	 \frac{2N_c}{\alpha_s} \int \frac{d^2 b_{1} d^2 b_{2}}{(2\pi)^4} e^{-i\Delta_\perp \cdot\frac{b_1+ b_2}{2} -ik_\perp\cdot(b_1-b_2)} \notag \\ 
% & \times \frac{1}{N_c}  \big \langle \text{tr}\big[(\partial_{\perp}^j U_{b_1}) (\partial_{\perp}^jU^\dagger_{b_{2}})\big]  \big \rangle_x~,
%    \label{eq:GTMD1}
%\end{align}

 The dipole-type gluon GTMD is defined as
    \begin{align}
        &x\mathcal{G}_{x,\textrm{DP}}(k_{\perp }, \Delta_\perp) = ~ 2\int \frac{\text{d}\xi^{-}\text{d}%
        ^{2}\xi _{\perp } e^{-ik_{\perp }\cdot \xi _{\perp }-ixP^{+}\xi ^{-}}}{\left(
        2\pi \right) ^{3}P^{+}} 
        \notag \\ 
        &\times \bigg\langle  P+\frac{ \Delta_\perp}{2}\bigg|\text{Tr}\bigg[ F^{+i}\left(\frac{\xi}{2}\right)
        \mathcal{U}_{[-\xi/2,\xi/2]}^{\left[ -\right] \dagger }\nn
        &\phantom{\times \bigg\langle  P+\frac{ \Delta_\perp}{2}\bigg|} \times F^{+i}\left( -\frac{\xi}{2}\right) \mathcal{U}_{[-\xi/2,\xi/2]}^{\left[ +%
        \right] }\bigg] \bigg| P-\frac{ \Delta_\perp}{2}\bigg\rangle\,, \label{eq:GTMD1}
    \end{align}
    where $\mathcal{U}^{[\pm]}$ are the future/past-pointing U-shaped Wilson lines, which make the operator gauge invariant. 
     The future/past-pointing U-shaped Wilson lines can be written as
\begin{align} 
    &\mathcal{U}_{[x,y]}^{\left[ \pm\right] }= U(x^-,\pm\infty;x_\perp)U(\pm\infty,y^-;y_\perp) \quad \nn
    &\text{with} \quad
    U(a, b; x_\perp) = \mathcal{P} \exp \left[ ig \int_a^b dx^- A_c^+(x^-, x_\perp) t^c \right],
\end{align}
where $\mathcal{P}$ represents the path ordered product. In the small \(x\) limit, this GTMD becomes
    \begin{align}
        x\mathcal{G}_{x, \textrm{DP}  }(k_{\perp }&, \Delta_\perp) = \frac{2N_{c}}{\alpha_s}\int
        \frac{d^{2}b_{1} d^{2}b_{2}}{(2\pi)^4} e^{ik_{\perp }\cdot \left(
        b_{1}-b_{2}\right) +i \frac{\Delta_\perp}{2}\cdot(b_{1}+b_{2})} \nn
        &\times\left(\nabla _{b_{1}}\cdot
        \nabla _{b_{2}}\right)\frac{1}{N_{c}}\left\langle\text{Tr}\left[
        U_{b_{1}} U^{\dagger }_{ b_{2}}
        \right]\right\rangle_x\,,
    \end{align}
where $U_{b_i}:=U(-\infty, \infty; b_i) $.%represents the Wilson line extending from $x^{-}=-\infty$ to $x^{-}=+\infty$ at the transverse position $b_i$, and $\langle \cdots\rangle_{x}$ stands for the CGC average.

   The WW-type gluon GTMD is defined by the following expression
    \begin{align}
        &x\mathcal{G}_{x, \textrm{WW}  }(k_{\perp }, \Delta_\perp) = ~ 2\int \frac{\text{d}\xi^{-}\text{d}%
        ^{2}\xi _{\perp } e^{-ik_{\perp }\cdot \xi _{\perp }-ixP^{+}\xi ^{-}}}{\left(
        2\pi \right) ^{3}P^{+}}
         \notag \\ 
        &\times  
        \bigg\langle  P+\frac{ \Delta_\perp}{2}\bigg|\text{Tr}\bigg[ F^{+i}\left(\frac{\xi}{2}\right)
        \mathcal{U}_{[-\xi/2,\xi/2]}^{\left[ +\right] \dagger }\nn
        & \phantom{\times \bigg\langle  P+\frac{ \Delta_\perp}{2}\bigg|} \times F^{+i}\left( -\frac{\xi}{2}\right) \mathcal{U}_{[-\xi/2,\xi/2]}^{\left[ +%
        \right] }\bigg] \bigg| P-\frac{ \Delta_\perp}{2}\bigg\rangle~.
    \end{align}
In the small \(x\) limit, it becomes
    \begin{align}
        x\mathcal{G}_{x,\textrm{WW}}(k_\perp&, \Delta_\perp) = \frac{2N_c}{\alpha_s} \int \frac{d^2 b_{1}d^2 b_{2}}{(2\pi)^4} e^{ik_\perp \cdot (b_{1} - b_{2}) + i \frac{\Delta_\perp}{2} \cdot (b_{1} + b_{2})} \nn
        &\times \frac{1}{N_c} \left\langle \text{Tr} \left[ i\partial_i U_{b_{1}} U^\dagger_{ b_{2}} i\partial_i U_{ b_{2}} U^\dagger_{b_{1}} \right] \right\rangle_x.
    \end{align}
Detailed derivations of the above relations can be found in Ref.~\cite{Dominguez:2011wm} and reference therein. %Similar to the QED case, the gluon A-GFF can be obtained by integrating the GTMDs over \( q_\perp \) and \( x \) as follows
Remarkably, after integrating out transverse momentum $k_\perp$ and using the identities $(\partial^j_\perp U_{b_\perp}) U^\dagger_{b_\perp}=-U_{b_\perp}(\partial^j_\perp U^\dagger_{b_\perp})$ and $U_{b_\perp} U^\dagger_{b_\perp}=1$, these two GTMDs reduce to the same gluon generalized parton distribution at zero skewness as well as the same gluon A-GFF. 

Finally, one can arrive at an interesting general relation between the gluon A-GFF and the Laplacian of the dipole scattering amplitude as follows
\begin{align}
A_g(t)
 = & \frac{2N_c}{\alpha_s} \int^1_0 d x\int \frac{ d^2  b_\perp}{(2\pi)^2} e^{-i \Delta_\perp \cdot  b_\perp }  \notag \\
 & \times  \vec \nabla_{ r_\perp}^2 \left[1-S_x( b_\perp, r_\perp)\right] \Big \vert_{ r_\perp= 0}~.
  \label{eq:Ag}
\end{align}
Here $S_x(b_\perp, r_\perp)=\frac{1}{N_c}\big\langle\operatorname{tr}\big[ U_{b_\perp+r_\perp / 2} U_{b_\perp-r_\perp / 2}^{\dagger}\big]\big\rangle_x$ represents the scattering amplitude between the color dipole with size $ r_\perp$ and the target hadron at impact parameter $ b_\perp$. In the CGC formalism, the dipole scattering amplitude usually follows the small-$x$ evolution governed by the non-linear small-$x$ evolution equation~\cite{Balitsky:1995ub,Jalilian-Marian:1997qno,Jalilian-Marian:1997jhx,Kovchegov:1999yj,Iancu:2000hn,Ferreiro:2001qy} with certain initial conditions~\cite{McLerran:1993ka,McLerran:1994vd,Golec-Biernat:1998zce,Kowalski:2003hm}. Also, phenomenological studies based on models and fitting of experimental data can provide useful information on $S_x(b_\perp, r_\perp)$~(see recent reviews~\cite{Frankfurt:2022jns,Mantysaari:2020axf}). Similarly, one can explore the relation to the BFKL intercept for $A_g(t)$ using the $b_\perp$-dependent BFKL solution given in e.g., \cite{Lipatov:1985uk,Navelet:1997xn,Hatta:2016dxp} and reference therein.

Eq.~(\ref{eq:Ag}) has a wide range of implications. First, if we set $\epsilon_g( b_\perp)=\frac{2N_c}{\alpha_s} \int^1_0 d x  \vec \nabla_{ r_\perp}^2  \left[1-S_x( b_\perp, r_\perp)\right] \big \vert_{ r_\perp= 0}$, we can interpret it as the gluon light-front momentum density, since it is related to the A-GFF via the Fourier transform $A_g(t)=\int \frac{ d^2  b_\perp}{(2\pi)^2} e^{-i \Delta_\perp \cdot  b_\perp }\epsilon_g( b_\perp) $. Therefore, the gluon transverse mean square radius~\cite{Freese:2021czn} is given by
\begin{align}
 \langle  b^2_\perp \rangle_g= \frac{\int_0^1 dx\int d^2  b_\perp b^2_\perp \vec \nabla_{ r_\perp}^2 S_x( b_\perp, r_\perp)  \vert_{ r_\perp = 0}}{\int_0^1 dx\int d^2  b_\perp\vec \nabla_{ r_\perp}^2 S_x( b_\perp, r_\perp)  \vert_{ r_\perp= 0}}~.
 \label{eq:bg}
\end{align}  
As expected, the above formula is equivalent to the result computed from the slope of the gluon A-GFF at $t=0$, as in the photon counterpart in Eq.~(\ref{Eq:def}). 

Second, the A-GFF of a QCD dipole (onium state), which is akin to the dipole photon A-GFF in Eq.~(\ref{eq:dipoleAGFF}), can be obtained by substituting $S_x(b_\perp, r_\perp)$ in Eq.~(\ref{eq:Ag}) with the onium-onium scattering amplitude~\cite{Kovchegov:2012mbw}.

Third, if one assumes the Gaussian form for the scattering amplitude $S_x( b_\perp, r_\perp) = \exp[ -\frac{ r^2_\perp}{4}Q_s^2(x,b_\perp)]$~\cite{Golec-Biernat:1998zce} with the saturation momentum $Q_s$ and neglects possible $\ln 1/r_\perp^2$ dependence when taking the Lapalacian, one can obtain the following A-GFF and radius:
 \begin{align}
 A_g(t)
 = &\frac{2N_c}{\alpha_s} \int^1_0 d x\int \frac{ d^2  b_\perp}{(2\pi)^2} e^{-i \Delta_\perp \cdot  b_\perp } Q_s^2(x,b_\perp)~,
 \notag \\ 
 \langle  b^2_\perp \rangle_g=& \frac{\int_0^1 dx\int d^2  b_\perp b^2_\perp Q_s^2( x, b_\perp)  }{\int_0^1 dx\int d^2  b_\perp  Q_s^2(x, b_\perp) }~.
 \label{eq:Qs}
\end{align} 
The physical meaning above results becomes manifest when we further adopt the Impact Parameter dependent Saturation (IP-Sat) model~\cite{Bartels:2002cj,Kowalski:2003hm} for proton. Then, the saturation momentum $Q_s(x, b_\perp)$ can be parametrized in the form
 \begin{align}
 Q_s^2(x,b_\perp)=\frac{2\pi^2 \alpha_s }{ N_c} xg(x, \mu^2 ) T( b_\perp)~.
 \label{eq:Qs2}
 \end{align} 
In the IP-Sat model, the $x$-dependence is given by the collinear gluon distribution $g(x, \mu^2)$ with the scale $\mu^2 =  \mu_0^2+4/r_\perp^2$ and the initial scale $\mu_0^2$ given by data fits, while the proton profile function $T( b_\perp)$ accounts for the impact parameter dependence. Thus, one finds the gluon A-GFF $A_g(t) = A_g(0) \int d^2b_\perp e^{-i \Delta_\perp \cdot b_\perp} T(b_\perp)$ with $A_g(0) = \int^1_0 dxxg(x, \mu^2) $. Since the dipole size $r_\perp$ is set to $0$ (or $1/Q$ with $Q$ the hard scattering scale) in Eq.~(\ref{eq:Ag}), one may take the asymptotic limit and choose $\mu^2 \simeq Q^2$, thus find the asymptotic gluon momentum fraction $\frac{4C_F}{4C_F+n_f}$~\cite{Gross:1974cs, Ji:1996ek} for $A_g(0)$ with the quark flavor number $n_f$. In principle, we can numerically check that the low-$x$ region with sufficiently large $\mu^2$ contributes significantly. In practice, one may use a phenomenological model, such as the IP-Sat model, which provides a smooth prescription of the scattering matrix across the low-$x$ to the large-$x$ region. The above results demonstrate the relationship between the gluon EMT form factor and the gluon light-front momentum radius~\cite{Freese:2021czn} $\langle  b^2_\perp\rangle_g=\int d^2  b_\perp b^2_\perp T( b_\perp)$ with the proton shape profile, which has been an important topic extensively studied in the small-$x$ formalism~(e.g.,~\cite{Kowalski:2003hm,Schlichting:2014ipa,Mantysaari:2016ykx,Mantysaari:2016jaz,Demirci:2022wuy}). 

Additionally, it will be interesting to study the impact of the small-$x$ evolution with impact-parameter dependence on $Q_s^2(x,b_\perp)$ and the related A-GFF. It should be noted that the small-$x$ evolution, such as that given by the BK equation, does not inherently lead to a finite radius~\cite{Kovner:2002yt,Kovner:2001bh}. Non-perturbative regularizations, particularly those related to QCD confinement, are necessary to ensure the finiteness of the gluon radius $\langle b_\perp^2 \rangle_g$~(see e.g.,~\cite{Berger:2010sh,Berger:2011ew} ), as assumed in phenomenological models such as IP-Sat. We leave it for a future work.

Furthermore, exclusive diffractive processes, such as the vector meson production and the deeply virtual Compton scattering in $ep$ scattering, provide gateways to constrain the proton density profile~(e.g.,~\cite{Kowalski:2006hc,Munier:2001nr,Rezaeian:2012ji,Toll:2012mb,Kowalski:2003hm,Caldwell:2010zza,Mantysaari:2018nng,Mantysaari:2018zdd, Boer:2023mip}), thus offer important inputs to the gluon A-GFF and the transverse radius through Eqs.~(\ref{eq:Ag}, \ref{eq:bg}). For instance, assuming a Gaussian profile $T( b_\perp)= \frac{1}{2\pi B_G} e^{-b^2_\perp/(2 B_G)}$~\cite{Kowalski:2003hm} in the IP-Sat model, one finds that the gluon radius and the A-GFF of the proton (insensitive to $\gamma^\ast p$ collisional energies and $x$ values) are given by $\sqrt{\langle  b^2_\perp \rangle_g}=\sqrt{2B_G}
$ and $ A_g(t)=A_g(0)e^{-B_G|t|/2}$, respectively. Based on the IP-Sat model, the fitting with the experimental data collected in exclusive diffractive processes at HERA leads to $B_G=4.0\pm 0.4~$GeV$^{-2}$~\cite{Rezaeian:2012ji}, resulting in $\sqrt{\langle  b^2_\perp \rangle_g}= 0.56 \pm 0.03$~fm. Exploring models~\cite{Iancu:2003ge,Marquet:2007nf,Marquet:2007qa,Watt:2007nr,Soyez:2007kg} akin to IP-Sat and investigating geometric fluctuations~\cite{Schlichting:2014ipa,Mantysaari:2016ykx,Mantysaari:2016jaz} and color charge correlations~\cite{Dumitru:2021tvw,Dumitru:2018vpr,Dumitru:2020fdh,Dumitru:2020gla} can offer further insights into gluon GFFs and radii.

In addition, in light of future EIC experiments, it is natural to extend the above analysis to high-energy electron-nucleus collisions. In particular, when the atomic mass number $A$ is sufficiently large, one can use the well-known McLerran-Venugopalan model~\cite{McLerran:1993ni,McLerran:1993ka,McLerran:1994vd} and compute the corresponding gluon A-GFF and radius. Since gluons interact with the down and up quarks in the same way, it is reasonable to expect that the gluon densities $g(x)$ inside the proton and neutron are largely identical. Thus, one can write the saturation momentum of the target nucleus as $Q_s^2(x,b_\perp)=\frac{2\pi^2 \alpha_s }{ N_c} xg(x) T_A( b_\perp)$ with the nuclear thickness function $T_A( b_\perp)$. If one simply supposes that the nucleon density is uniform inside a large spherical nucleus with the geometric radius $R_A$~\cite{Mueller:1999yb, Mueller:1999wm,Kowalski:2003hm}, then one finds $T_A( b_\perp)=\frac{3A}{2\pi R^3_A}\sqrt{R^2_A- b^2_\perp}$. This simple profile function then leads to $\langle  b^2_\perp \rangle_g=\frac{2}{5}R^2_A=\frac{2}{3}\langle  r^2 \rangle_g$ with the three-dimensional radius $\langle  r^2 \rangle$ and $A_g(t)=A \int_0^1 dx xg(x) f_N(t)$ with $f_N(t)=3\frac{\sin  R_A\Delta_\perp - R_A \Delta_\perp\cos R_A\Delta_\perp}{R_A^3 \Delta_\perp^3}$. More sophisticated theoretical inputs and EIC measurements, combined together, allow us to precisely pinpoint the nucleon distribution inside nuclei including properties like density fluctuations~\cite{Mantysaari:2020axf}. The WW method used in this work offers an interesting perspective at high energy, complementing previous studies of nuclear GFFs~\cite{Polyakov:2002yz,Guzey:2005ba,Liuti:2005qj,GarciaMartin-Caro:2023klo,GarciaMartin-Caro:2023toa}.

Last but not least, by studying the gluon mean square radius, one can even indirectly extract the neutron distribution information in medium and large nuclei. High energy electron-nucleus collisions at EIC allow us to probe the proton and neutron densities collectively, while measurements of the electromagnetic and weak charge form factors are separately sensitive to the proton and neutron distributions, respectively. Let us write the nuclear profile function $T_A(b_\perp)$ as the sum of the transverse density distributions $\rho(b_\perp)$ for protons and neutrons:
\begin{align}
 T_A(b_\perp)=\rho_p(b_\perp)+ \rho_{n}(b_\perp)~.
 \end{align}
Thus, the transverse mean square radii of proton and neutron distributions can be cast into
 \begin{align}
 \langle  b^2_\perp \rangle_{p,n}= \frac{\int d^2 b_\perp b_\perp^2 \rho_{p,n}(b_\perp)}{\int d^2 b_\perp \rho_{p,n}(b_\perp)}~.
 \end{align} 
Eventually, one can derive a simple equation relating these radii to the nuclear gluon radius as follows
\begin{align}
 \langle  b^2_\perp \rangle_{g}= \frac{Z}{A}\langle  b^2_\perp \rangle_{p}+\frac{A-Z}{A}\langle  b^2_\perp \rangle_{n}~.
 \label{eq:SR_radius}
 \end{align} 
Here, the symbols $A$ and $Z$ represent the atomic mass and atomic numbers, respectively. Experimentally, the three radii in Eq.~(\ref{eq:SR_radius}) can be measured separately. The nuclear gluon radius, which is related to the nucleon distribution in a nucleus, can be constrained by exclusive processes like diffractive vector meson production~(see~e.g.,~\cite{Caldwell:2010zza,Lappi:2010dd,Mantysaari:2020axf,Mantysaari:2023qsq}). Meanwhile, the proton mean square radius can be obtained from the charge form factor~\cite{Miller:2010nz} $\langle  b^2_\perp \rangle_{p}=\frac{4}{F_{\text{ch}}(0)}\frac{dF_{\text{ch}}(t)}{dt}\Big \vert _{t=0}$ in elastic electron scatterings. In addition, as the weak charge of the neutron is about $14$ times of that of the proton, the measurement of the weak charge form factor and the parity violating effect in electron scattering can probe the neutron distribution (see~\cite{Mammei:2023kdf} and references therein). Additionally, the mean square transverse radii in Eq.~(\ref{eq:SR_radius}) are directly related to the conventional 3D radii through the following relation: $\langle b^2_\perp \rangle_{n,p}=\frac{2}{3}\langle r^2\rangle_{n,p} $~\cite{Miller:2009sg} for a large nucleus with spherical symmetry. In nuclear physics, the neutron skin thickness~\cite{Thiel:2019tkm,Mammei:2023kdf}, defined as $S=\langle r^2\rangle_{n}^{1/2}-\langle r^2\rangle_{p}^{1/2}$, is a central focus in the studies of neutron-rich matter~(see~\cite{PREX:2021umo,Reed:2021nqk,Giacalone:2023cet} for $^{208}$Pb and the recent STAR measurement\cite{STAR:2022wfe}). Therefore, using Eq.~(\ref{eq:SR_radius}), EIC can open a gateway to constraining the neutron skin from the nuclear gluon radius and supplement to the direct measurements through the parity violating electron scatterings.

\section{Summary } In summary, we have applied the WW method to compute the photon and gluon A-GFFs of fast-moving hadrons and have established a simple relation between the gluon A-GFF and the color-dipole scattering amplitude. With this relation, we can extract the gluon transverse radius and explore important information about the energy-momentum distribution of gluons inside hadrons. Furthermore, we suggest that the nuclear gluon radius can be written as the sum of the charge and weak charge radii of nuclei, providing a useful constraint on the neutron distributions. Our study represents only the initial step towards understanding the small-$x$ effects in the gluon GFFs and corresponding radii with the eikonal approximation in the high-energy limit. It provides analytical insights and complementary information when compared to lattice simulations and experimental measurements. By incorporating sub-eikonal corrections using the theoretical framework developed in Refs.~\cite{Altinoluk:2014oxa,Altinoluk:2015gia,Altinoluk:2020oyd,Kovchegov:2015pbl,Kovchegov:2017lsr,Kovchegov:2018znm,Chirilli:2018kkw}, we expect that the extension to other GFFs may become possible. With the advent of the upcoming EIC, we can gain a deeper understanding of the internal structure of hadrons and the related emergent phenomena.

{\bf Acknowledgments:} 
We thank Alfred Mueller, Feng Yuan, Yoshitaka Hatta, Tuomas Lappi, Heikki Mäntysaari, Zhuoyi Peng, Yi Chen, Yi Yin, and Yuxiang Zhao for useful conversations. This work is supported in part by the CUHK-Shenzhen university development fund under grant No. UDF01001859. X.B. is supported by the Research Council of Finland, the Centre of Excellence in Quark Matter and supported under the European Union’s Horizon 2020 research and innovation programme by the European Research Council (ERC, grant agreements No. ERC-2023-101123801 GlueSatLight and No. ERC-2018-ADG-835105 YoctoLHC) and by the STRONG-2020 project (grant agreement No. 824093). The content of this article does not reflect the official opinion of the European Union and responsibility for the information and views expressed therein lies entirely with the authors.

\end{document}